\documentclass[runningheads]{llncs}

\usepackage{url}
\usepackage{graphicx}
\usepackage{booktabs}
\usepackage{makecell}
\usepackage{listings}
\usepackage{color}
\usepackage{enumitem}

\usepackage{hyperref}
\hypersetup{
    colorlinks=true,
    linkcolor=black,
    citecolor=black,
    filecolor=black,
    urlcolor=black,
}

\newcommand{\ignore}[1]{}

\begin{document}
\title{The Performance Envelope of Inverted Indexing on Modern Hardware}

\authorrunning{Lin et al.}

\author{Jimmy Lin, Lori Paniak, and Gordon Boerke}

\institute{David R. Cheriton School of Computer Science\\
University of Waterloo}

\maketitle

\begin{abstract}
This paper explores the performance envelope of ``traditional'' inverted indexing on modern hardware using the implementation in the open-source Lucene search library.
We benchmark indexing throughput on a single high-end multi-core commodity server in a number of configurations varying the media of the source collection and target index, examining a network-attached store, a direct-attached disk array, and an SSD.
Experiments show that the largest determinants of performance are the physical characteristics of the source and target media, and that physically isolating the two yields the highest indexing throughput.
Results suggest that current indexing techniques have reached physical device limits, and that further algorithmic improvements in performance are unlikely without rethinking the inverted indexing pipeline in light of observed bottlenecks.


\end{abstract}

\section{Introduction}

Despite the advent of multi-stage reranking architectures and neural models for document ranking, the humble inverted index remains an indispensable data structure for end-to-end information access applications from document retrieval to question answering.
Despite alternative approaches based on approximate nearest-neighbor search~\cite{Boytsov_etal_CIKM2016,Malkov2019}, inverted indexes---in combination with modern query evaluation algorithms such as block-max \textsc{Wand}~\cite{Ding_Suel_SIGIR2011}---remain the standard by which other retrieval techniques are judged.
In this paper, we focus on inverted indexing applied to static document collections and explicitly leave aside the so-called real-time indexing problem, where high velocity document streams need to be ingested and made immediately searchable; such a scenario calls for different techniques than when working with static collections~\cite{Busch_etal_ICDE2012,Asadi_etal_SIGKDD2013,ZhangDongxiang_etal_TOIS2017a}.

While the 1990s and 2000s saw much research activity in ``traditional'' inverted indexing on static document collections~\cite{Moffat_Bell_JASIS1995,Heinz_Zobel_JASIST2003,Lester_etal_CIKM2005,Buttcher_etal_SIGIR2006b,Lester_etal_2008}, there has been relatively few innovations in the last decade or so (leaving aside quasi-succinct indexes, which belong in a separate family of techniques altogether).
It appears that inverted indexing is now considered ``mature technology''.
Putting this assertion to the test, we benchmark the open-source Lucene search engine on modern hardware, with the aim of characterizing the performance envelope of the world's most widely-deployed implementation.
We examine different sources from which we read the document collection as well as different targets into which we write the index structures, testing a network-attached store, a direct-attached disk array, and an SSD.
We find that the largest determinants of performance are the physical characteristics of the source and target media, and that isolating the two yields the highest indexing throughput.

The contributions of this work are as follows:\
We conducted a thorough experimental evaluation of the performance of a production inverted indexing implementation on a modern server, exploring a range of hardware configurations.
Results suggest that current indexing implementations have reached physical device limits.
We discuss the implications of these findings, arguing that further {\it algorithmic} contributions to performance are unlikely without rethinking the inverted indexing pipeline.

\section{Methods}

At a fundamental level, inverted indexing involves the transformation of a document collection held in stable storage into index structures that also reside in stable storage at the end of the process.
Throughout this paper, we refer to the source (where the document collection resides, where we are reading from) and the target (where the indexes are ultimately held, where we are indexing into).
Popular classes of stable storage include network-attached stores, direct-attached disk arrays, and SSDs.

In this paper, all our experiments were conducted on a server with two Intel Xeon Platinum 8160 processors (33M Cache, 2.10 GHz, 24 cores each) with 1 TB RAM, running Ubuntu 18.04.
This machine was purchased in mid-2019 and can be characterized as ``high end'', but firmly in the commodity class.
We specifically explore the following storage configurations:

\smallskip \noindent 
{\bf Ceph.}
We made use of a Ceph storage cluster primarily as read-only network-attached storage for holding the document collection off server.
The Ceph cluster is configured as a warm-storage redundant system with three servers, each with 36 6TB 7200RPM hard drives.
An underlying ZFS file system on each server guarantees data fidelity and implements compression.
A Ceph file system (CephFS) is constructed on the Ceph cluster to store and provide access to the source collections.
Finally, this CephFS is translated to NFS using the Ganesha VFS translator.
It is this NFS export that is mounted for use on the working server.
The cluster is connected to our working server via 10 GbE links.
Note that while CephFS (imperfectly) supports POSIX semantics, we use Ceph primarily as a read-only store of the source collection.

\smallskip \noindent 
{\bf Direct-attached disk array.}
Local storage on our server consists of 24 8TB 7200RPM hard drives.
Two configurations were tested with the local storage array:\ ZFS and AVAGO 3108 MegaRAID/XFS.
Each has its own strengths and different regimes of suitability.
Our configurations were based on hard drive topologies with data striped across four six-drive sets, where each set has two parity/redundant drives.

\begin{itemize}
    
\item The ZFS configuration is a 4-vdev RAIDZ2 pool, where we store data in a ZFS file system with the following options: \texttt{atime=off}, \texttt{acltype=posixacl}, \texttt{xattr=sa}, \texttt{.compression=on}.

\item The AVAGO configuration is a RAID60. The AVAGO RAID volume is formatted with an XFS file system with default settings.

\end{itemize}

\noindent The XFS configuration provides higher file system read and write performance over ZFS.
This is expected due to the overhead in the aggressive data integrity mechanisms in ZFS.
A Merkle tree of checksums guarantees the validity of each block of data in the ZFS array at write time and when the block is read.
While these precautions protect data against bit-rot, phantom writes, and various storage subsystem errors, they require more CPU resources.

\smallskip \noindent {\bf SSD.}
In addition to the disk array described above, we have a Samsung SM883 3.84TB SATA SSD formatted to \texttt{ext4} with default settings.
The SSD is connected to the system via the Intel SATA controller.

\smallskip \noindent
Given these storage media, a number of source--destination pairs makes sense:
\begin{itemize}
\item From Ceph into the direct-attached disk array.
\item From Ceph into the SSD.
\item From the direct-attached disk array into the direct-attached disk array.
\item From the direct-attached disk array into the SSD.
\item From the SSD into the SSD.
\end{itemize}
\noindent For the direct-attached disk array, we experimented with both ZFS and XFS on the exact same machine:\ we ran all experiments with ZFS and then rebuilt the file system from scratch with XFS.
Thus, our ZFS/XFS experiments are directly comparable.
To reduce the number of experimental conditions we discarded some source--target combinations that do not make sense.
For example, our use of Ceph does not make it suitable as an indexing target; also, we see no compelling case to indexing from SSDs into spinning disks.

Our indexing experiments used the ClueWeb09b (CW09b) and ClueWeb12-B13 (CW12b) web crawls, provided by Carnegie Mellon University; CW09b comprises 50.2M pages totaling 231GB compressed while CW12b comprises 52.3M pages totaling 389GB.


Indexing was performed with the open-source Anserini toolkit~\cite{Yang_etal_JDIQ2018} (v0.6.0),\footnote{\url{http://anserini.io/}} as Lucene itself provides no support for working with the two web collections.
Our version of Anserini is based on Lucene 8.0, which is the latest major Lucene release.
We use Oracle Java 11.0.2.
We built full positional indexes and store the parsed document vectors as well as a copy of the original raw documents alongside the inverted index.
This choice best simulates a real-world search application:\
The parsed document vectors support efficient relevance feedback and downstream reranking, and of course, users ultimately want to view documents; the original distribution of the ClueWeb collections does not support efficient random access.
Thus, once the index has been built, the original collection is no longer needed.

Each experimental condition was run three times on the otherwise idle working server, with a physical system reboot before each trial to ensure that nothing was kept resident in caches.
All experiments were run with 48 threads to maximize use of all available cores.
In an alternate set of experiments, we tried running with 96 threads to take advantage of hyper-threading.
The results were not substantively different (sometimes a bit faster, sometimes a bit slower), and thus these results are not reported for space considerations.

\section{Results}

\begin{table}[t]
\centering\resizebox{\textwidth}{!}{
\begin{tabular}{lll lrr lrr}
\toprule
& & & \multicolumn{3}{c}{{\bf CW09b} (50.2M pages, 231GB)} & \multicolumn{3}{c}{{\bf CW12b} (52.3M pages, 389GB)} \\
\cmidrule(lr){4-6}  \cmidrule(lr){7-9}  
& & & time & GB/m & docs/s & time & GB/m & docs/s  \\
\midrule
Ceph & $\rightarrow$ & ZFS & 2:27:12 $\pm97$  & 1.57 & 5.69 $\times$ 10$^3$ & 2:56:12 $\pm27$  & 2.21 & 4.95 $\times$ 10$^3$\\
ZFS  & $\rightarrow$ & ZFS & 2:28:29 $\pm1$   & 1.56 & 5.64 $\times$ 10$^3$ & 2:58:41 $\pm78$  & 2.18 & 4.88 $\times$ 10$^3$\\
\midrule
Ceph & $\rightarrow$ & XFS & 1:33:19 $\pm35$  & 2.48 & 8.97 $\times$ 10$^3$ & 1:51:31 $\pm25$  & 3.49 & 7.82 $\times$ 10$^3$\\
XFS  & $\rightarrow$ & XFS & 1:56:30 $\pm19$  & 1.98 & 7.18 $\times$ 10$^3$ & 3:06:04 $\pm127$ & 2.09 & 4.69 $\times$ 10$^3$\\
\midrule
Ceph & $\rightarrow$ & SSD & 0:59:30 $\pm88$  & 3.88 & 1.41 $\times$ 10$^4$ & 1:19:39 $\pm143$ & 4.88 & 1.10 $\times$ 10$^4$\\
ZFS  & $\rightarrow$ & SSD & 1:14:14 $\pm54$  & 3.11 & 1.13 $\times$ 10$^4$ & 1:37:24 $\pm134$ & 3.99 & 8.96 $\times$ 10$^3$\\
XFS  & $\rightarrow$ & SSD & 0:57:37 $\pm56$  & 4.01 & 1.45 $\times$ 10$^4$ & 1:15:42 $\pm64$  & 5.14 & 1.15 $\times$ 10$^4$\\
SSD  & $\rightarrow$ & SSD & 1:28:23 $\pm127$ & 2.61 & 9.47 $\times$ 10$^3$ & 1:57:14 $\pm271$ & 3.32 & 7.44 $\times$ 10$^3$\\
\bottomrule
\end{tabular}
}
\vspace{0.1cm}
\caption{Indexing performance, with time reported in h:mm:ss $\pm$ standard deviation (in seconds) and throughput reported in raw compressed gigabytes per minute and documents per second.}
\label{table:results}
\end{table} 

Our experimental results are shown in Table~\ref{table:results}, where each row represents a different source--target combination, grouped by target (more below).
Columns are grouped by collection:\ CW09b and CW12b.
For each, we report indexing time in h:mm:ss with $\pm$ standard deviation in seconds.
Performance is additionally characterized in terms of gigabytes (of raw compressed collection) indexed per minute and documents indexed per second.
Note that while CW09b and CW12b have roughly the same number of documents, CW12b is larger due to the growth of webpage size; for the most part, this is JavaScript and other non-indexable material that is stripped in the document processing pipeline.

For reference, the complete CW09b index is 685GB and the complete CW12b index is 869GB.
Note that the complete index is larger than the raw collection because, along with the full positional indexes, we also store the parsed document vectors and the raw documents (as noted above).
Although Lucene stores the documents in compressed form, because of the requirement for random access, this compression is not as efficient as in the raw collection.

At a high level, Lucene indexing performance is quite impressive---in the best configuration, it is able to achieve an indexing throughput of hundreds of gigabytes per hour and tens of thousands of documents per second.
We see, however, quite big performance differences across different hardware configurations, {\it with exactly the same code}.
Indexing CW09b can take as little as an hour or as long as 2.5 hours; CW12b indexing time ranges from 1.3 hours to slightly over 3 hours.
In both cases, the maximum difference is roughly a factor of three.
It appears that physical characteristics of the media matter the most to performance, likely more so than the impact of different indexing implementations.


With a bit of analysis, we find write throughput to be the current performance bottleneck---during indexing, we observe consistent write throughput of $\sim$500MB into the SSD, which bumps up against the maximum throughput of the SATA controller.
Thus, when indexing into the SSD, the source (Ceph, ZFS, XFS) has less impact on performance.
It appears that Ceph and XFS can fully ``keep up'' with the SSD writes, while ZFS lags a bit, and thus is a bit slower.
Indexing from the SSD into the SSD, as expected, is much slower than reading the raw collection from another media because the controller needs to split its bandwidth between reads and writes.
Thus, for optimal performance, it makes sense to isolate the source media from which the indexer is reading the raw documents and the target media that the index is being built on.
Since the bottleneck appears to be on SSD writes, reading source documents across the network (Ceph) does not appear to be significantly slower than reading documents from local storage (XFS).
This is an interesting observation because the network {\it does not} appear to be the bottleneck in this case, even with 10 GbE links.
The growing popularity of 40 or even 100 GbE links gives us further performance headroom.

Even on the direct-attached disk array, the choice of file system appears to make a big difference---compare the rows involving XFS and ZFS.
If we consider the case where we are isolating the reads (i.e., indexing from Ceph), XFS is approximately 40\% faster than ZFS.
Once again, to emphasize:\ we are running exactly the same code, on exactly the same hardware.
These performance differences are purely the result of the design of ZFS and XFS.
The better data protection offered by ZFS is very costly from the performance perspective.

\section{Discussion}

On immediate and important lesson for future researchers is that it is {\it absolutely critical} to accurately document the file system configuration when reporting the results of large-scale indexing experiments.
We observed differences up to a factor of three with different media configurations, which is a huge experimental variable that needs to be properly isolated and studied when proposing and evaluating new indexing techniques.

While our experiments only benchmarked one particular inverted indexing implementation, we believe that it is possible to generalize from our findings and discuss implications for future work.
However, there are two important points about Lucene that are worth emphasizing:

First, Lucene is the most widely-deployed solution for building production search applications in the world, used by a long list of organizations that  include Apple, Bloomberg, Reddit, Twitter, and Wikipedia.
For many practitioners, Lucene is synonymous with search; most practitioners have never heard of academic systems that are commonly used in research papers.
Thus, a better understanding of Lucene's indexing performance is alone of great value to the broader community.


Second, Lucene's inverted indexing implementation is actually very well engineered and captures many of the lessons learned from the academic literature; for example, Lucene 8 introduced block-max indexes, which is fairly close to the state of the art, even from the academic perspective.
At a high level, the indexing pipeline is based on in-memory inversion with periodic flushes to disk; partial index segments are then merged in a hierarchical manner.
A consequential design decision is to have each indexing thread operate on its own set of documents, writing to its own index segment.
This minimizes coordination overhead and exploits modern multi-core processors, but the design choice means that the index segments are relatively small (compared to the alternative of having multiple concurrent writers), which places pressure on downstream index merges.

The pipeline analogy is quite apt to conceptualize the current state of affairs.
We can visualize inverted indexing as a pipe connecting a source to a sink.
Our experiments show that SSD write throughput forms the current bottleneck.
Informally, the end of the pipe is too narrow.
That is, the SSD can't keep up with all the writes (flushes) from the cores performing in-memory inversion, although in some cases (for example, ZFS $\rightarrow$ SSD), we find evidence suggesting that the source media can't supply documents fast enough (i.e., limits on read bandwidth or the source end of the pipe).
Current technological trends mean that the problem is going to become increasingly worse as core counts continue to grow.
In our analogy, the ``middle'' of the pipe is becoming fatter and fatter.
Since we scale out indexing by having each core work independently on its own set of documents, pressure on downstream index merges (writes into the target media) will continue to increase as well.
Yet, any alternative to independent indexing threads requires heavyweight concurrency coordination, which means that the cores would not be working at their full potential.

We are able to draw general conclusions here because the limitations discussed above are conceptual, not implementation-specific.
While more careful software engineering can help, the fundamental calculus does not change---for example, better compression can reduce write pressure, but core counts are likely increasing faster than improvements in compression.
This means that current inverted indexing techniques have reached device limitations---for example, we could likely further improve performance by building SSD arrays and exploiting multiple controllers to increase bandwidth, but these can be characterized as brute force hardware-based solutions.
It appears that substantial {\it algorithmic} improvements in performance are not possible without rethinking the entire inverted indexing pipeline.

\section{Conclusion}

This paper contributes to our understanding of inverted indexing in two main ways:\
For the practitioner, we provide some useful advice when trying to optimize indexing throughput.
The choice of physical media matters a lot, and it is a good idea to isolate the source and target.
For the researcher, we provide some ``food for thought'' that hopefully will inspire future work:\ optimal system performance is achieved when all components are in balance, but the trend of increasing core counts means that I/O performance (be it network-attached storage, direct-attached disk arrays, or SSDs) are fast becoming the bottleneck.
Overcoming this bottleneck, we believe, requires fundamentally rethinking the entire inverted indexing pipeline.

\section*{Acknowledgments}

This research was funded by the Natural Sciences and Engineering Research Council (NSERC) of Canada, with additional support from Start Smart Labs and the Global Water Futures program.

\bibliographystyle{splncs04}

\end{document}